\documentclass[reprint,aps,prd,superscriptaddress,showkeys]{revtex4-1}
\usepackage{epsfig,amsmath,natbib}

\usepackage{aas_macros}
\usepackage{amssymb}
 \usepackage{amsmath}
 \usepackage{dsfont}
 \usepackage{hyperref}
 \usepackage{color}
\hypersetup{
	colorlinks=false,
	citecolor=green
}

\begin{document}

\title{The impact of spurious shear on cosmological parameter estimates from weak lensing observables}

\author{Andrea Petri}
\email{apetri@phys.columbia.edu}
\affiliation{Department of Physics, Columbia University, New York, NY 10027, USA}
\affiliation{Physics Department, Brookhaven National Laboratory, Upton, NY 11973, USA}

\author{Morgan May}
\affiliation{Physics Department, Brookhaven National Laboratory, Upton, NY 11973, USA}

\author{Zolt\'an Haiman}
\affiliation{Department of Astronomy, Columbia University, New York, NY 10027, USA}

\author{Jan M. Kratochvil}
\affiliation{Astrophysics and Cosmology Research Unit, University of KwaZulu-Natal, Westville, Durban 4000, South Africa}

\date{\today}

\label{firstpage}

\begin{abstract}
Residual errors in shear measurements, after corrections for
instrument systematics and atmospheric effects, can impact
cosmological parameters derived from weak lensing observations.  Here
we combine convergence maps from our suite of ray-tracing
simulations with random realizations of spurious shear. This allows us to
quantify the errors and biases of the triplet $(\Omega_m,w,\sigma_8)$
derived from the power spectrum (PS), as well as from three different
sets of non-Gaussian statistics of the lensing convergence field:
Minkowski functionals (MF), low--order moments (LM), and peak counts
(PK). Our main results are: (i) We find an order of magnitude smaller
biases from the PS than in previous work. (ii) The PS and LM yield
biases much smaller than the morphological statistics (MF, PK).  (iii) For strictly Gaussian spurious shear with integrated amplitude as low as its current estimate of $\sigma^2_{sys}\approx 10^{-7}$, biases from the PS and LM would be unimportant even for a survey with the statistical power of LSST.  However, we find that for surveys larger than $\approx 100$ deg$^2$, non-Gaussianity in the noise (not included in our analysis) will likely be important and must be quantified to assess the biases. (iv) The morphological statistics (MF,PK) introduce important biases even for
Gaussian noise, which must be corrected in large surveys. The biases
are in different directions in $(\Omega_m,w,\sigma_8)$ parameter
space, allowing self-calibration by combining multiple statistics.
Our results warrant follow-up studies with more extensive lensing
simulations and more accurate spurious shear estimates.
\end{abstract}

\keywords{Weak Gravitational Lensing --- Data analysis --- Methods: analytical,numerical,statistical}

\maketitle


\section{Introduction}

Weak gravitational lensing (WL) offers one of the most promising
cosmological probes (see pedagogical reviews by
\cite{BartelmannSchneider,RefReview, HoekstraJain2008}, as well as a
recent review of WL in the context of other cosmology probes
\cite{Weinberg+2013}).  Results from the first large observational
surveys, such as COSMOS~\cite{Schrabback2010} and
CFHTLenS~\cite{Kilbinger2013}, obtained interesting constraints with the technique, and found constraints generally
compatible with the accepted $\Lambda$CDM cosmology. Ongoing surveys such as the Dark Energy Survey (DES) \citep{DESMain} and Hyper Suprime-Cam  (HSC) \citep{HSCMain}, and future surveys such as LSST \citep{LSSTMain} and Euclid \citep{EuclidMain} will greatly improve these constraints.

Most of the attention to date has focused on utilizing the power
spectrum of the cosmic shear (or equivalent two-point functions, such
as the angular correlation function).  However, in recent years, the
strongly non-Gaussian nature of the WL disortion field on small
($\sim$arcminute) angular scales have been increasingly considered.
Non-Gaussian features can, in principle, allow more information to be
extracted from the same datasets, using higher-order statistics.  The
proposed beyond-Gaussian statistics have included
the one-point function (e.g. \cite{JSW00,SW09}),
the bispectrum (e.g. \cite{T&J04,BergeAmaraRefregier10}),
skewness and higher moments (e.g. \cite{Hui99,T&J02,Petri2013}),
shapelets and flexions (e.g. \cite{Refregier2003,GoldbergBacon2005}),
the abundance \cite{PeaksJan,DietrichHartlap10,MarianBernstein06,PeaksXiuyuan}
and clustering \cite{Marian+13} of peaks, 
and Minkowski functionals
\cite{Guimaraes02,Maturi+09,MinkJan,Munshi12,Shirasaki+2012,Petri2013,Connolly2012}.

Many of these higher-order statistics have recently been detected in
large WL datasets.  The third-order moments of the aperture mass have
been measured in COSMOS and found to modestly tighten constraints on
$\Omega$ and $\sigma_8$ \cite{Semboloni+2011}.  In the CFHTLenS
survey, moments up to fourth order \cite{VanWaerbeke+13}, the number
counts and correlation functions of peaks \cite{Shanetal+14},
three-point correlations functions \cite{Fu+2014}, and Minkowski
functionals \cite{ShirasakiYoshida2014} have all been measured. The
latter two statistics have also been shown to tighten cosmological
constraints compared to using two-point statistics alone.


The recent progress in utilizing non-Gaussian statistics motivates us
to study the impact of systematic errors on these statistics.  A vast
body of work exist on the impact and mitigation of systematic errors
for the PS and the correlation functions
(e.g. \cite{Huterer2006,Bartelmann+2012}, but the analogous effort has
not yet been made to compute the impact of the same errors on
non-Gaussian statistics. Exceptions include recent studies which
considered the effect of uncorrelated galaxy shape, instrumental and atmospheric measurement errors
on peak counts \cite{DebbiePeaks}, and the impact of masking
\cite{Shirasaki+2013}, as well as photo-$z$ errors and additive and
multiplicative shear errors \cite{ShirasakiYoshida2014} on the MFs.
In this paper, we extend these previous works, and we study the
confidence limits and the biases of cosmological parameters, arising
from residual systematic errors in shear measurements. These errors (herafter referred as``spurious shear''),
which are left after correcting for the point spread function (PSF), are small, but are correlated between different
directions on the sky. In general, there is no reason to expect the
spurious shear to obey Gaussian statistics.

This paper is a first step towards quantifying the impact of spurious
shear on four sets of weak lensing statistics PS, MF, LM, and PK, in a
uniform fashion, using a restriced suite of ray-tracing simulations
(the ``Inspector Gadget Suite \#1'', hereafter IGS1). In this first
study, another major simplifying assumption, which we will relax in
future work, is that the spurious shear is Gaussian.  Thus, the goal
of the present paper is to quantify the impact of noise whose power
spectrum is different from that of random, uncorrelated shape noise.

The rest of this paper is organized as follows. 
In \S~\ref{formalism}, we give an overview of the statistical
formalism we use to compute the parameter constraints and biases from
each observable. 
In \S~\ref{sysnoise}, we discuss the log-linear model spurious shear power spectrum that based on published estimates by \citep{amara}.
Our results are then presented in \S~\ref{results}. We first use
lensing power spectra computed semi-analytically with the public code
NICAEA \citep{nicaea} to validate our simulations and to compare
errors and biases from the PS to previous work.  We then use the
simulations to compute errors and biases from the non-Gaussian
statistics (LM,MF,PK).
In \S~\ref{lsstnoise}, analyze the properties and effects of spurious shear which \citep{chihway} derived from a detailed simulation of the LSST instrument, focusing on non-Gaussianities and deviations from the published power
spectrum on small angular scales.
In \S~\ref{discussion}, we discuss our results and their limitations
more generally, and propose several future improvements.
Finally, in \S~\ref{conclusions}, we summarize our conclusions and the
implications of this work.

\section{Formalism}
\label{formalism}

\subsection{Observables}
In this section we give a brief overview of the cosmological probes we use to calculate the constraints on the $\Lambda$CDM cosmological parameters. Weak lensing probes are based on the idea that, given a galaxy at redshift $z$ (or equivalently at comoving distance $\chi(z)$), the dark matter density fluctuations $\delta$ between that galaxy and observers on Earth will generate distortions in the observed galaxy shape. These shape distortions are parametrized by the convergence $\kappa$, which is related to the magnification, and by the two components of shear $(\gamma_1,\gamma_2)$, which are related to the ellipticity. As already widely suggested in the literature (see for example \citep{BartelmannSchneider}), we can probe the convergence field $\kappa$ measuring its power spectrum $P_l$, which is directly related to the power spectrum of the fluctuations in the 3D gravitational potential $\Phi(\mathbf{x})$
\begin{equation}
\label{crosspower}
P_l^{z_1z_2}=\frac{\pi^2l}{2}\int_0^{\infty}dz \frac{d\chi}{dz}\frac{g^{z_1}(z)g^{z_2}(z)}{\chi^3}P_\Phi\left(\frac{l}{\chi},z\right).
\end{equation} 
Here $P_\Phi(k,z)$ is the gravitational potential power spectrum and $g^{z_i}(z)$ is a redshift weight function that depends on the redshift distribution of galaxies in the redshift bin $z_i$; the power spectrum is a quadratic descriptor of the convergence field $\kappa$ and hence one can consider single redshift correlators $P_l^z$ or double redshift correlators $P_l^{z_1z_2}$ (which reduce to single redshift correlators if $z_1=z_2$), which are the ones that are used in the first part of the analysis. Equation (\ref{crosspower}) tells us that $P_l$ is essentially a projection of the gravitational potential power spectrum along the line of sight. Additional cosmological probes for the convergence field $\kappa$ that we consider in this paper are all real space statistics, namely a particular class of low--order moments (LM), Minkowski functionals (MF) and peaks (PK). 
These additional probes might be particularly useful for constraining cosmological parameters since the $\kappa$ field is heavily non-Gaussian, and will contain information beyond the power spectrum (see for example \citep{DietrichHartlap10,MarianBernstein06,BergeAmaraRefregier10} for some proposed methods of extracting non-Gaussian information from weak lensing data). 
The moments we consider consist of the set of the two quadratic moments $\sigma_0^2=\langle\kappa^2\rangle$ and $\sigma_1^2=\langle\vert\nabla\kappa\vert^2\rangle$, three cubic moments $(S_0,S_1,S_2)$ and four connected quartic moments $(K_0,K_1,K_2,K_3)$, see \citep{Munshi12,Petri2013}. Minkowski functionals $(V_0(\nu),V_1(\nu),V_2(\nu))$ are topological descriptors of the convergence field: $V_0$ is related to the area of the excursion set $\{\kappa>\nu\sigma_0\}$, $V_1$ to the length of its boundary and $V_2$ to its genus characteristic (see \citep{Tomita90,Tomita}). Perturbative expansions of the Minkowski functionals in terms of the moments of the convergence have been studied in \citep{Matsubara10,Munshi12}, but have been shown not to converge with sufficient accuracy in \citep{Petri2013}. Finally, the peak statistic $N(\nu)$ counts the number of local maxima of amplitude $\nu\sigma_0$ in the convergence field. The efficiency of this statistic in constraining cosmology has been studied in \citep{PeaksXiuyuan,PeaksJan}. In this work, we study the effect of spurious shear on the constraints obtained from these cosmological probes.  

\subsection{Power spectrum: lensing tomography formalism}
To compare with previous work and check our simulations, we use the public code NICAEA \citep{nicaea} to compute the $\kappa$ (convergence) cross power spectrum $P_l^{z_iz_j}$. If we restrict ourselves to the quasi-Gaussian $l$ modes (typically $l\lesssim$ few$\times 10^3$), 
we have a good model for the power spectrum covariance matrix, and we can build a Fisher matrix (see, e.g. \cite{FangHaiman07}),
\begin{equation}
\label{fishertomo}
F_{\alpha\beta} = \frac{f_{sky}}{2}\sum_{z_{1,2,3,4}}\sum_{l=l_{min}}^{l_{max}}(2l+1)P_{l,\alpha}^{z_1z_2}W_l^{z_2z_3}P_{l,\beta}^{z_3z_4}W_l^{z_4z_1}
\end{equation}
that can be then used to compute marginalized constraints for the parameters
\begin{equation}
\label{errtomo}
e(p_\alpha) = \sqrt{(F^{-1})_{\alpha\alpha}}.
\end{equation}
Here $P_{l,\alpha}^{z_iz_j}$ is the derivative of the power spectrum with respect to the cosmological parameter $p_\alpha$, $f_{sky}$ is the fraction of sky covered by the survey and $W_l$ is the inverse of the power spectrum covariance matrix
\begin{equation}
W_l^{z_iz_j}=\left(P_l^{z_iz_j} + N_l^{z_iz_j} + S_l^{z_iz_j}\right)^{-1}.
\end{equation}
In eq.~(\ref{fishertomo}), and throughout the rest of this paper, we adopt the Einstein summation convention over repeated indices (i.e. there is an implied summation over redshift bins).
The inverse has to be calculated with respect to the redshift indices. 
Here we consider three main contributions to the covariance matrix, namely the signal itself ($P_l$), the galaxy shape noise ($N_l$) and an additional source of spurious shear due to the instrument and the atmosphere ($S_l$). We model the galaxy shape noise as a redshift-dependent, uncorrelated white noise component $N_l^{z_iz_j}=N_{z_i}\delta^{z_iz_j}$, with amplitude (see \citep{SongKnox})
\begin{equation}
N_z = \frac{(0.15+0.035z)^2}{n(z)}
\end{equation} 
where $n(z)$ is the galaxy density per unit redshift per unit solid angle. One can compute the total galaxy density per unit solid angle as 
\begin{equation}
n_g=\int_{z_{min}}^{z_{max}}\frac{dn(z)}{dz}dz.
\end{equation}
We use the normalization $n_g=30\,\mathrm{arcmin}^{-2}$ (see below).
The model for the spurious shear $S_l$ is instrument--dependent, and will be discussed in \S~\ref{sysnoise}. For the moment we will consider a redshift--independent
spurious shear power spectrum $S_l^{z_iz_j}=S_l\delta^{z_iz_j}$. 
If ignored, the spurious shear can introduce a bias $b(p_\alpha)$ in the cosmological parameters, which can be quantified as 
\begin{equation}
\label{biastomo}
b(p_\alpha) = \frac{f_{sky}}{2}F_{\alpha\beta}^{-1}\sum_{z_{1,2,3}}\sum_{l=l_{min}}^{l_{max}}S_lW_l^{z_1z_2}W_l^{z_2z_3}P_{l,\beta}^{z_3z_1}
\end{equation}
This is the tomographic generalization of the method used in \citep{amara} to compute the parameter biases for a single--redshift galaxy sample. In this work we compare the marginalized errors $e$, to the biases $b$ to see in which conditions the latter are important, under the assumption that the spurious effects are purely Gaussian, are redshift--independent, and are well described by a power spectral density $S_l$.  

\subsection{Beyond the power spectrum: the nonlinear statistics}
In this section, we describe the formalism to go beyond the power spectrum and calculate the marginalized errors and the biases on the cosmological parameters using the nonlinear statistics measured from the IGS1 simulations. 
Unlike the analytical power spectrum calculations, we do not have a good theoretical model 
for either the expectation values of our nonlinear observables (LM, MF, PK) or for their covariance matrix,
so we are forced to measure it from the simulations. 
The cosmological N-body simulations of large-scale structures and
ray-traced weak lensing maps used in this paper are the same as those
in our earlier work \cite{MinkJan, PeaksXiuyuan, Petri2013}.  We refer the reader to these
publications for a full description of our methodology; here we
review the main features.

A total of 80 CDM-only N-body runs were made with the IGS1
lensing simulation pipeline. Our suite of 7 cosmological models includes a fiducial model with
parameters \{$\Omega_m=0.26$, $\Omega_\Lambda=0.74$, $w=-1.0$,
$n_s=0.96$, $\sigma_8=0.798$, $h=0.72\}$, as well as six other
models.  In each of these six models, we varied one parameter at a
time, keeping all other parameters fixed at their fiducial values; we
thus have WL maps in variants of our fiducial cosmology with
$w=\{-0.8, -1.2\}$, $\sigma_8=\{0.75, 0.85\}$, and $\Omega_m=\{0.23,
0.29\}$.  Note that in the last case, we set $\Omega_\Lambda=\{0.77,
0.71\}$ to keep the universe spatially flat.

To produce the N-body simulations, we first created linear
matter power spectra for the seven different cosmological models with
CAMB \citep{CAMB}
for $z=0$, and scaled them back to the starting
redshift of our N-body simulations at $z=100$ following the linear
growth factor.  Using these power spectra to create initial particle
positions, the N-body simulations were run with a modified version of
the public N-body code GADGET-2 \citep{Gadget-2} 
and its accompanying
initial conditions generator N-GenIC.  We modified both codes to allow
the dark energy equation of state parameter to differ from its
$\Lambda$CDM value ($w\neq 1$), as well as to compute WL-related
quantities, such as comoving distances to the observer, at each
simulation cube output. Each simulation contains $512^3$ CDM particles
in a box size of $(240h^{-1}{\rm comoving~Mpc})^3$, allowing a mass
resolution of $7.4\times10^9h^{-1}M_\odot$.

In each of the six non-fiducial cosmological models, we ran 5 strictly
independent N-body simulations (i.e. each with a different realization
of the initial conditions). To minimize the differences between two
cosmologies arising from different random realizations, the initial
conditions for each of those five simulations were matched across the
cosmologies quasi-identically.  This entails recycling the same random
number when drawing mass density modes from the power spectrum for
each cosmology (note that the power spectra themselves of course
differ across the cosmologies).  In the fiducial cosmology, we ran 50
strictly independent simulations -- the first set of 5 to match the
other cosmologies quasi-identically as mentioned above, and an
additional set of 45 to improve the statistical accuracy of the
predictions in the fiducial cosmology (especially the covariance
matrices).  In each cosmology we generated 1000 pseudo-independent 12 deg$^2$ maps of $\kappa,\gamma_1$ and $\gamma_2$ using the ray--tracing algorithm in \citep{HamanaMellier}; in this paper we focus on the $\kappa$ maps. 

Measurements of observables from the simulations require binning for both the power spectrum and the topological statistics. The finite size of the maps we use, $\theta_{map}$ forces power spectral modes which differ by less than $2\pi/\theta_{map}$ to be grouped in the same $l$ bin, and the continuous nature of the $V_0(\nu),V_1(\nu),V_2(\nu)$ (MF) and $N(\nu)$ (PK) statistics requires the threshold $\nu$ to be discretized in finite intervals in order to make the Fisher matrix calculations tractable. The effects of binning choices  on our results is investigated in \S~\ref{robustness} (see also \citep{Petri2013} for reference). We refer as $O_i^r$ to the set of observables measured in each realization $r$ of the fiducial cosmology, where the index $i$ can range from 1 to the number of bins $N_{bins}$ chosen for the PS, MF and PK statistics, and from 1 to 9 for the LM statistic (to include all nine moments, no binning is required for the LM statistic). Given the IGS1 ensemble of $R=1000$ realizations, we can measure the average and covariance matrix of the observables
\begin{equation}
\langle O_i\rangle=\frac{1}{R}\sum_{r=1}^RO_i^r,
\end{equation}
and
\begin{equation}
\label{covest}
C_{ij}=\frac{1}{R-1}\sum_{r=1}^R(O_i^r-\langle O_i\rangle)(O_j^r-\langle O_j\rangle).
\end{equation}
We use the non--fiducial simulated maps to measure the finite--difference derivatives of the observables vector $X_{i\alpha}\equiv \langle O_i\rangle,_\alpha=\partial \langle O_i\rangle/\partial p_\alpha$. We fit each realization for the cosmological parameters $p_\alpha$ using a $\chi^2$ minimization as in \citep{Petri2013}
\begin{eqnarray}
\nonumber
\delta p^r_\alpha = p^r_\alpha - p^0_{\alpha} &=& (X_iC^{-1}_{ij}X_j)^{-1}_{\alpha\beta}(X_{k,\beta}C^{-1}_{kl})(O^r_l - \langle O_l\rangle)\\
&=&M_{\alpha l}   (O^r_l - \langle O_l\rangle),
\label{parest}
\end{eqnarray}
where $p_\alpha^0$ are the fiducial cosmological parameters and $M_{\alpha l}$ is a shorthand for 
\begin{equation}
\label{mat}
M_{\alpha l}\equiv (X_iC^{-1}_{ij}X_j)^{-1}_{\alpha\beta}X_{k\beta}C^{-1}_{kl}.
\end{equation}
Note that the IGS1 simulations are limited to variation of the triplet $p_\alpha=(\Omega_m,w,\sigma_8)$.

If the average $\langle O_i\rangle$ and derivatives $X$ that we use to build the model are computed using the maps without spurious shear (i.e. with just galaxy shape noise added), then the estimator (\ref{parest}) is biased, and the amount of bias (in the small bias limit) is given by 
\begin{equation}
\label{bias}
b(p_\alpha)=M_{\alpha i}(\langle O^S_i\rangle - \langle O_i\rangle),
\end{equation}
and $O^S$ are the observables calculated from the simulations with spurious shear included.
We can also quantify the parameter covariance matrix 
\begin{equation}
\label{parcov}
P_{\alpha\beta} = \langle\delta p_\alpha^r\delta p_\beta^r\rangle = M_{\alpha i}M_{\beta j}C^{S}_{ij}
\end{equation}
where $C^S$ is the covariance matrix of the observables calculated with spurious shear effects included. We can then calculate the marginalized parameter constraints as 
\begin{equation}
\label{marginalized}
e(p_\alpha)=\sqrt{P_{\alpha\alpha}}\,\,.
\end{equation} One should note that, if we set $C^S=C$, equation (\ref{parcov}) reduces to the usual Fisher matrix expression. 


\section{Modeling Spurious Shear}
\label{sysnoise}
In this section, we give a description of how we model the spurious shear that contaminates the actual lensing signal. We distinguish two types of additive systematic errors:  uncertainties due to the shape measurement technique, and atmospheric and instrumental effects which can distort the recorded galaxy images. Additive effects due to measurement techniques are usually modelled as a white noise source (see \citep{Huterer2006}) and hence almost indistinguishable from intrinsic galaxy shape noise; the net effect of this kind of additive systematic is to decrease the effective galaxy number density of the survey. Reference \citep{DebbiePeaks} considered the effect of uncorrelated shape measurement, instrumental and atmoshperic errors on shear peak statistics for LSST. In this work, we concentrate on the correlated atmospheric and instrumental (mainly due to optics) effects, which we call \textit{spurious shear}, following \citep{chihway}. Unlike shape measurement errors, spurious shear is correlated between pixels. Its power spectral shape has been fitted by a log-linear model (see \citep{chihway,amara}). We perform our analysis on convergence maps, i.e. on the $E$ mode of the shear, and we model the spurious shear correlations by means of a power spectral density of the form
\begin{equation}
\label{logpower}
S_{EE,l}\equiv S_l=\frac{A}{l(l+1)}\left\vert1+n\log\left(\frac{l}{l_0}\right)\right\vert
\end{equation}  
with $l_0=700$ and $(A,n)$ kept as adjustable parameters. For simplicity, we restrict ourselves to the case where the spurious shear is purely Gaussian, and is fully characterized by the power spectrum of equation (\ref{logpower}). This assumption will likely have an important effect on the results we obtain using the nonlinear statistics (LM,MF,PK) which are particularly sensitive to non-Gaussianities. We discuss this issue further in \S~\ref{lsstnoise} below.

\begin{figure}
\begin{center}
\includegraphics[scale=0.45]{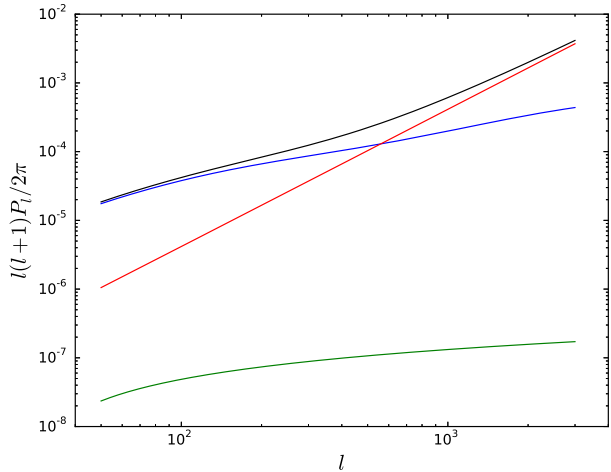}
\end{center}
\caption{One of the diagonal components of the convergence cross power spectrum $P_l^{zz}$ with $z=1.95$; the plot displays pure signal $P_l$ (blue), signal shape noise $N_l$ (red) added using a total galaxy density $n_{gal}=30\mathrm{arcmin}^{-2}$ (which fixes the normalization in equation (\ref{galaxydistr})), and spurious shear $S_l$ (green) with spectral index $n=0.7$, with normalization $\sigma_{sys}^2=4\times10^{-7}$. For reference we also plotted the total convergence cross spectrum in black}
\label{powerspectra}
\end{figure}

 To have a sense of the orders of magnitude, we display in Figure \ref{powerspectra} the power spectra calculated with NICAEA, separating the signal contribution from those of shape noise and spurious shear. In \S~\ref{lsstnoise} we find that the log-linear model is not a good description of the actual LSST simulated atmospheric maps on scales smaller that $\sim 3^\prime$. For the moment, we will ignore this complication; we will investigate the effects of this small--scale departure from the log-linear model in \S~\ref{lsstnoise} below.
Figure \ref{powerspectra} shows that, for an integrated spurious shear power spectrum with $\sigma^2_{sys}=4\times 10^{-7}$ (the expression for $\sigma^2_{sys}$ can be found in equation (\ref{integratedsys}), and this particular value has been chosen according to \citep{amara,chihway}), the power in spurious shear is much smaller than the shape noise. 

For the simulation analysis of the nonlinear statistics, we generate $R=1000$ noise maps in Fourier--space, which are random realizations of the power spectrum of equation (\ref{logpower}), and FFT invert them (using the FFTW3 C library \citep{FFTW05}) to create the real space noise maps that we add to the simulated convergence maps. 


\section{Results}
\label{results}

\subsection{Analytical results with NICAEA}

The purpose of this subsection is to illustrate some analytical results we obtained using the power spectrum code NICAEA: the goal is to compare these analytical results to the full numerical ones, which can give an estimate of the IGS1 simulations' accuracy. We begin by computing parameter errors and biases semi-analytically, so that we can compare our 
results with previously published work.

We consider a flat $\Lambda$CDM model with 7 cosmological parameters, which can have the two alternative parametrizations 
\begin{equation}
\label{models}
\begin{matrix}
p^1_\alpha=(\Omega_mh^2,\Omega_{DE},w_0,w_1,\sigma_8,\Omega_bh^2,n_s) \\
p^2_\alpha=(\Omega_m,w_0,w_1,h,\sigma_8,\Omega_b,n_s).
\end{matrix}
\end{equation}
The former is easier to deal with when including Planck priors in the analysis, while the latter is necessary to make a comparison with \citep{amara}. 

We consider a galaxy distribution with exponential tails as in \citep{amara,SongKnox}
\begin{equation}
\label{galaxydistr}
n(z)\propto z^\alpha\exp{\left[-\left(\frac{z}{z_0}\right)^\beta\right]}
\end{equation}
%
%
in which the peak redshift scale is $z_0$ and the parameters $\alpha,\beta$ are chosen to match observations. We divide the galaxies into redshift bins $z_i$ and use NICAEA to compute the convergence cross power spectrum $P_l^{z_iz_j}$; we subsequently apply equations (\ref{fishertomo}) -- (\ref{biastomo}) to compute the biases and marginalized constraints on the cosmological parameters. 

\subsubsection{Comparison with LSST figure of merit}
Before proceeding to calculate the cosmological constraints and biases, we investigated the importance of the choice of the step $\delta p_\alpha$ used to calculate the derivatives $P_{l,\alpha}^{z_iz_j}$, since in general the errors and biases will depend on it. We developed an iterative method in which we start choosing an initial step (20\% variations in the parameters), compute the $1\sigma$ marginalized errors with equation (\ref{errtomo}) and use these errors as the new steps to compute $P_{l,\alpha}^{z_iz_j}$. This ensures that the final constraints we obtain are based on models for which the observables were actually computed in the Fisher derivatives. In practice we find that only a few iterations are necessary in order to get convergent results. A sample of our marginalized constraints after 30 iterations can be found in Table \ref{bias7parTable} on the top. 

\begin{figure}
\begin{center}
\includegraphics[scale=0.4]{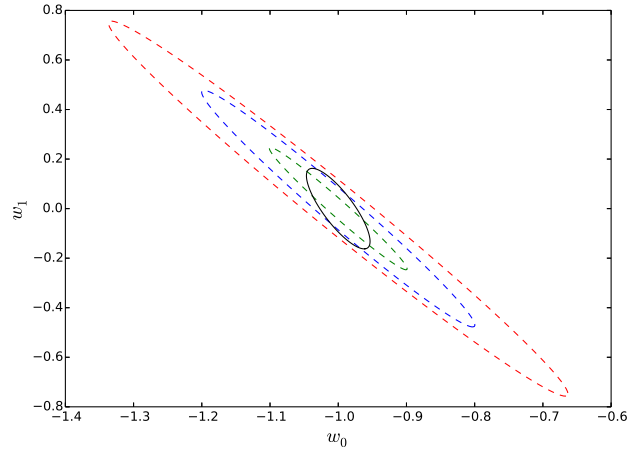}
\end{center}
\caption{Comparison between our error calculations on the doublet $(w_0,w_1)$ (solid black ellipse) with a $5\%$ step used for calculating the derivatives, and the LSST figure of merit (dashed ellipses) with no systematics (green), optimistic additive systematics (blue) and pessimistic additive systematics (red) as published in \citep{LSST2.0}. Planck priors $\Delta_\alpha$ were added, resulting in an effective Fisher matrix $F^{\Delta}_{\alpha\beta}=F_{\alpha\beta}+\Delta^{-1}_{\alpha}\delta_{\alpha\beta}$}
\label{lsstmerit}
\end{figure}

A comparison with the LSST figure of merit for the $(w_0,w_1)$ doublet is displayed in Figure \ref{lsstmerit} (in order to make this comparison, we have added Planck priors, see \citep{PlanckXVI2013} for reference). The survey specifications used for the comparison are the same as in Table \ref{bias7parTable} top, and we limited ourselves to the case in which $S_l\equiv 0$ (no spurious shear). The figure shows that our results agree with the published systematic--free curve within a factor of $\sim 2$.  

\subsubsection{Bias estimation}
In this section we study the effect of introducing a log-linear spurious shear as in equation (\ref{logpower}), with an integrated amplitude of $\sigma_{sys}^2=4\times 10^{-7}$ defined as 
\begin{equation}
\label{integratedsys}
\sigma^2_{sys}=\frac{1}{2\pi}\sum_{l=l_{min}}^{l_{max}}lS_l.
\end{equation}
We proceed as in \citep{amara} switching to the $p_\alpha^2$ parametrization of equation (\ref{models}) and setting the survey specifications to the ones found in their paper; a complete overview of the survey assumptions and results we obtained with this framework can be found in Table \ref{bias7parTable}. 
\begin{table*}
\begin{tabular}{cccccccc}

\multicolumn{8}{c}{\textbf{Survey assumptions 1}} \\
\multicolumn{8}{c}{$(\alpha,\beta,z_0,f_{sky},n_g)=(2,1,0.7,0.35,30\,\mathrm{arcmin}^{-2})$, $z\in[0,3]$, $l\in[50,3000]$} \\ \hline

& $n_s$ & $w_1$ & $w_0$ & $\Omega_bh^2$ & $\sigma_8$ & $\Omega_{DE}$ & $\Omega_mh^2$ \\ \hline
Fiducial & 1.0 & 0.0 & -1.0 & 0.021 & 0.9 & 0.7 & 0.147 \\
Error & 0.020 & 0.17 & 0.043 & 0.0009 & 0.0059 & 0.0032 & 0.0090 \\ \hline \hline \\

\multicolumn{8}{c}{\textbf{Survey assumptions 2}} \\
\multicolumn{8}{c}{$(\alpha,\beta,z_0,f_{sky},n_g)=(2,1.5,0.64,0.44,35\,\mathrm{arcmin}^{-2})$, $z\in[0,4]$, $l\in[10,20000]$} \\ \hline

& $\Omega_m$ & $w_0$ & $w_1$ & $h$ & $\sigma_8$ & $\Omega_b$ & $n_s$ \\ \hline
Fiducial & 0.28 & -0.95 & 0.0 & 0.7 & 1.0 & 0.046 & 1.0 \\
Bias & 4.4$\times 10^{-5}$ & 6.5$\times 10^{-4}$ & $-$2.7$\times 10^{-3}$ & $-$4.1$\times 10^{-5}$ & $-$5.0$\times 10^{-5}$ & 1.1$\times 10^{-4}$ & 2.1$\times 10^{-4}$ \\
Error & 0.0087 & 0.14 & 0.59 & 0.12 & 0.011 & 0.019 & 0.034 \\
\end{tabular}
\caption{Fiducial values used for the parametrizations $p^{(1)}_\alpha$ (top)  and $p^{(2)}_\alpha$ (bottom), along with marginalized errors (top and bottom) and biases (bottom) calculated with equations (\ref{errtomo},\ref{biastomo}), for a sample value of $n=0.7$.}
\label{bias7parTable}
\end{table*}
It is worth noticing that the biases in Table \ref{bias7parTable} are about an order of magnitude smaller than the ones that \citep{amara} found. The origin of this factor of 10 difference is unclear. However, we note that \citep{amara} quotes unphysically large marginalized errors on $h$ ($\Delta h=17$) and $\Omega_b$ ($\Delta \Omega_b=4$) in their Table~1. While the origin of these large errors is also unclear, we suspect it may be related to the larger values of the biases they find. The computation of the bias involves manipulating the same matrices as those in computing the marginalized errors, and the large $h$ and $\Omega_b$ errors could increase the biases in other parameters through degeneracies.

\subsection{Comparison between NICAEA and simulations}
\label{compsection}
The purpose of this subsection is to compare the analytical results obtained with the NICAEA to the numerical results obtained from the simulations. 

\subsubsection{Bias and error comparison}

We compare the parameter constraints obtained from the convergence power spectrum (PS), using both the simulated maps and NICAEA. For this comparison, we focus on $l$ modes between $500\leq l \leq 5000$, with a bin step $\delta l_{bin}=100$ (roughly corresponding to the $l$ resolution of the maps $\delta l_{pix}=2\pi/\theta_{map}$), consider a single source plane at $z_s=2$, and we do not add any spurious shear (just galaxy shape noise). Since in this section we analyze the power spectrum statistic only, we do not smooth the maps, because the conclusions for the PS statistic are independent of smoothing scale. We then apply equations (\ref{mat})-(\ref{marginalized}), with $O_l=P_l$, and the covariance matrix approximated as diagonal 
\begin{equation}
\label{covdiag}
C^{th}_{ll^\prime}=\frac{P_l^2}{l+1/2}\delta_{ll^\prime}.
\end{equation}
Note that with this choice of covariance matrix equations (\ref{mat}) and (\ref{marginalized}) are the same as equations (\ref{biastomo}) and (\ref{errtomo}) in the limit of no tomography, with only one redshift bin and $f_{sky}=1$. Note also that this expression for the covariance matrix is correct only when we have full sky coverage, so that for each mode $l$ there are $2l+1$ identically distributed submodes. In a more realistic case, when we consider finite patches of sky of size $\theta_{map}$, we are limited by the size of the pixel in Fourier space $\delta l_{pix}=2\pi/\theta_{map}=\sqrt{\pi/f_{sky}}$. The size of the bins we use to probe the power spectrum, which we call $\delta l_{bin}$ must be comparable with this number. This gives us a number of submodes $N_{sub}(l)$, for each $l$ mode, with
\begin{equation}
N_{sub}(l)\approx\frac{\pi l \delta l_{bin}}{\delta l_{pix}^2}
\end{equation}
which will result in a measured covariance matrix
\begin{equation}
\label{covmeas}
C^{meas}_{ll^\prime}=\frac{P^2_l}{N_{sub}(l)}\delta_{ll^\prime}\approx\frac{\delta l_{pix}^2}{\pi\delta l_{bin}}\frac{P^2_l}{l}\delta_{ll^\prime}.
\end{equation} 
The results of this comparison are shown in Table \ref{comptable}, and they are scaled in a way that takes into account the fact that 
\begin{equation}
C^{meas}_{ll^\prime}\approx \frac{\delta l_{pix}^2}{\pi\delta l_{bin}}C^{th}_{ll^\prime}.
\end{equation}
\begin{table*}
\begin{center}
\begin{tabular}{ccccc}

\multicolumn{5}{c}{\textbf{Survey Assumptions 3}} \\
\multicolumn{5}{c}{$n(z)=n_g\delta(z-2)$, $n_g\gg 1$, $l\in[500,5000]$} \\ \hline

& $\Omega_m$ & $w$ & $\sigma_8$ & $\sigma_8\Omega_m^{0.5}$ \\ \hline
\multicolumn{5}{c}{\textbf{NICAEA}} \\ \hline
$b(p_\alpha)$ & $2.46\times 10^{-5}$  & $-2.29\times 10^{-3}$ & $8.57\times 10^{-6}$ & $2.37\times 10^{-5}$\\ 
$e(p_\alpha)$ & 0.035 & 0.10 & 0.055 & 0.0021\\ \hline
\multicolumn{5}{c}{\textbf{Simulations}} \\ \hline
 $b(p_\alpha)$ & $2.3\times 10^{-5}$ & $-1.3\times 10^{-4}$ & $1.23\times 10^{-5}$ & $2.43\times 10^{-5}$\\
 $e(p_\alpha)$ & 0.023 & 0.11  & 0.036 & 0.0028 \\ \hline
\end{tabular}
\end{center}
\caption{Bias and marginalized errors comparison on the parameters using the power spectrum computed with the code NICAEA and the IGS1 simulations. In the calculations with NICAEA we used 45 linearly spaced modes between $500\leq l \leq 5000$, while in the simulations we chose 45 linearly spaced $l$ bins in the same interval. The maps were not smoothed and a single source plane at redshift $z=2$ was considered. For simplicity, no galaxy shape noise was added and no Gaussian smoothing was applied to the maps}
\label{comptable}
\end{table*}
The results obtained with NICAEA and the simulations are in good agreement, especially when we mitigate the degeneracy between $\Omega_m$ and $\sigma_8$ by considering the parameter combination $\sigma_8\Omega_m^{0.5}$.  The discrepancy in the $w$ bias is large (a factor of $\sim 20$), and might be related to the inaccuracies in the $w$ derivative that we see in Figure \ref{derivatives} at high $l$ which we will discuss in the next section. Because of degeneracies between parameters, an inaccuracy in even one of the power spectrum derivatives can affect the constraints on the remaining parameters. However, we note that the $w$ bias is very small, well below the errors from any foreaseable Weak Lensing experiment.  

\subsubsection{Simulation inaccuracies and degeneracies}
\begin{figure}[htdp]
\begin{center}
\includegraphics[scale=0.4]{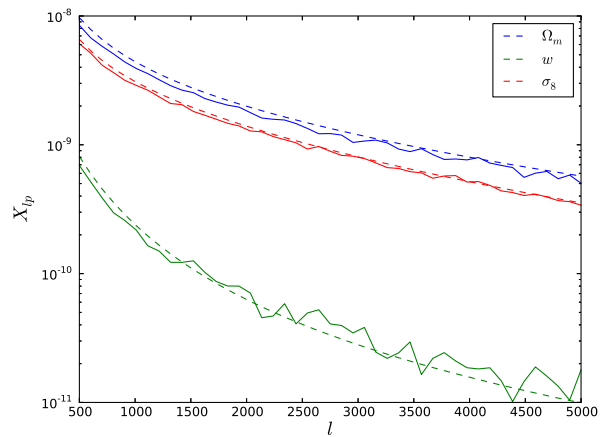}
\caption{Power spectrum derivatives measured from the simulations (solid) and from NICAEA (dashed) for the three parameters $\Omega_m$ (blue), $w$ (green) and $\sigma_8$(red). The agreement is good overall, but the noise visible on the plot causes significant inaccuracy in the calculated bias for $w$.}
\label{derivatives}
\end{center}
\end{figure}

Confidence intervals for the parameters $(\Omega_m,w,\sigma_8)$ obtained from NICAEA and from the simulations are shown in Table \ref{comptable}. The differences between the simulations and NICAEA could be due to an inaccuracy in measuring the derivatives $X$ from the simulations, or to noise in the simulated covariance matrix. We test these two hypotheses and conclude that the main source of the difference is the inaccuracy in measuring the power spectrum derivatives.
Figure \ref{derivatives}  shows a comparison between the power spectrum derivatives calculated with NICAEA and measured from the IGS1 simulations; we also quantified the numerical fluctuations in the covariance matrix measuring the dimensionless quantity $\langle\sqrt{l}\Delta P_l/P_l\rangle$ which, according to the Gaussian prediction should have an expectation value over $l$ of order of $\sqrt{\pi\delta l_{bin}}/\delta l_{pix}\approx 6$. We found our prediction to be consistent with the measurements up to $\sim 10\%$ random numerical noise when spacing multipoles $l\in [500,5000]$.
We calculate biases and parameter constraints by inserting into  equations (\ref{mat})-(\ref{parcov}), alternately the covariance matrices $C$ and the derivatives tensors $X$ measured from the simulations and those calculated semi-analytically with NICAEA. We find that choosing one particular method of measuring the covariance matrix has a negligible effect on the parameter constraints, whereas switching the analytically calculated derivatives with those inferred from the simulations has a large effect.

These inaccuracies have an important effect especially in the presence of degeneracies, like the well known $(\Omega_m,\sigma_8)$ one. 
To limit the effects of this degeneracy, instead of constraining both $\Omega_m$ and $\sigma_8$, we constrain the combination $\sigma_8\Omega_m^\gamma$. We choose $\gamma$  such that the variation $\delta(\sigma_8\Omega_m^\gamma)$ given by
\begin{equation}
\delta(\sigma_8\Omega_m^\gamma) = \Omega_m^\gamma \delta\sigma_8 + \gamma\sigma_8\Omega_m^{\gamma-1}\delta\Omega_m
\end{equation}
corresponds to the minor axis of the $(\Omega_m,\sigma_8)$ likelihood confidence contour (which has the shape of an ellipse, within our framework). We identify the directions of the ellipse axes by computing the eigenvectors of the parameter covariance matrix:  we find that regardless of whether we use the power spectra from NICAEA or the ones measured from our simulations, we obtain the same value $\gamma=0.5$. We display the constraints on the $\sigma_8\Omega_m^{0.5}$ combination as a fourth column in Table \ref{comptable}, which shows good agreement between the NICAEA and IGS1 errror contours.
\subsection{Results from simulations: the nonlinear statistics}
\label{simresults}
\begin{figure*}[htdp]
\begin{center}
\includegraphics[scale=0.3]{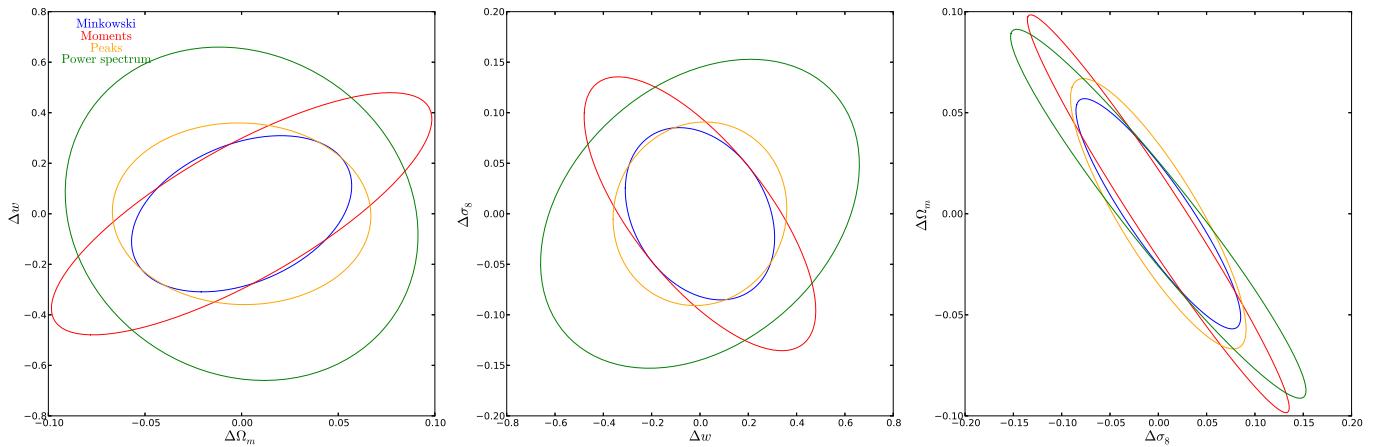}
\caption{Confidence contours corresponding to Table \ref{errbiastable}.}
\label{allcontours}
\end{center}
\end{figure*}

\begin{figure}[htdp]
\begin{center}
\includegraphics[scale=0.4]{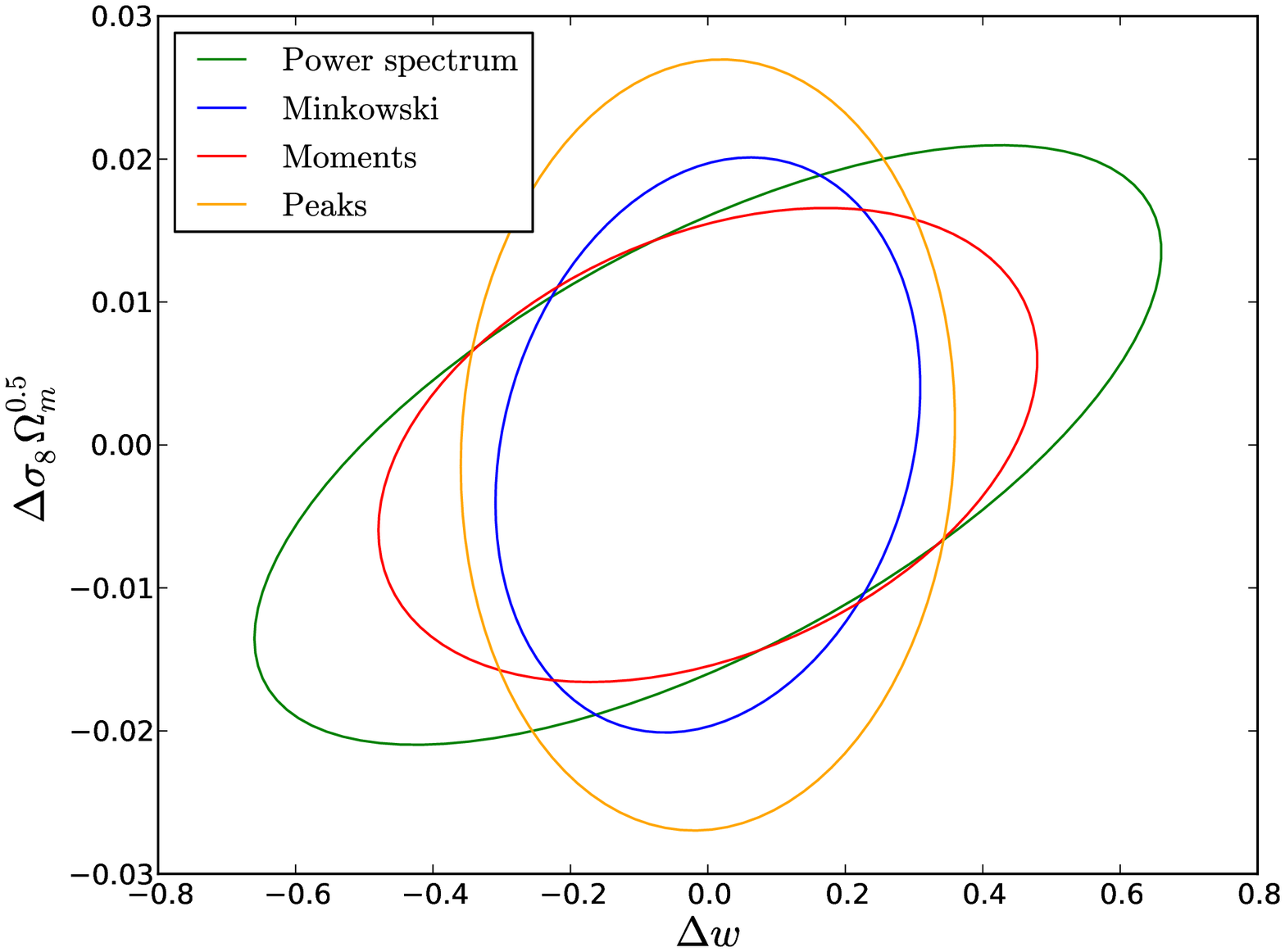}
\caption{Error contours in the $(w,\sigma_8\Omega_m^{0.5})$ plane}
\label{degcontours}
\end{center}
\end{figure}

We use the IGS1 set of simulated maps to evaluate the biases and marginalized errors (through equations (\ref{bias}) and (\ref{marginalized})) on the parameter triplet $(\Omega_m,w,\sigma_8)$ using, in addition to the power spectrum (PS), the three nonlinear statistics (LM,MF,PK). The results are outlined in Table \ref{errbiastable} and Figures \ref{allcontours} and \ref{degcontours}. One of the main conclusions that we draw from our results is that the marginalized constraints are comparable for all the different statistics, although the best among the nonlinear statistics, the MFs, deliver constraints that are a factor of $\sim 2$ better than those from the PS. We also note that, even though the PS appears to be the less constraining statistic, it also appears to have a much smaller bias than the morphological statistics (MF,PK). Also note that the LM statistic is much less biased than the topological ones. The main reason for this is that the spurious shear considered introduces large scale correlations that can affect the topology of the excursion set, but has virtually no effect on higher-than-quadratic point statistics, such as the skewness $S_0=\langle\kappa^3\rangle$.     
\begin{table*}
\begin{center}
\begin{tabular}{ccccc}

\multicolumn{5}{c}{\textbf{Survey Assumptions 4}} \\
\multicolumn{5}{c}{$n(z)=n_g\delta(z-2)$, $n_g=15\,\mathrm{arcmin}^{-2}$, $l\in[100,2\times10^4]$, $\nu^{MF}\in[-2,2]$, $\nu^{pk}\in[-2,5]$} \\ \hline

& $\Omega_m$ & $w$ & $\sigma_8$ & $\sigma_8\Omega_m^{0.5}$ \\ \hline
\multicolumn{5}{c}{\textbf{Power spectrum}} \\ \hline
$b(p_\alpha)$ & $4.0\times 10^{-6}$  & $-2.69\times 10^{-4}$ & $2.5\times 10^{-5}$ & $1.5\times 10^{-5}$ \\ 
$e(p_\alpha)$ & 0.060 & 0.43 & 0.10 & 0.014 \\ \hline
\multicolumn{5}{c}{\textbf{Minkowski}} \\ \hline
 $b(p_\alpha)$ &0.0026 &0.037 & $-0.0024$ & $8.31\times 10^{-4}$ \\
 $e(p_\alpha)$ &0.038 &0.20  &0.056 & 0.013\\ \hline
\multicolumn{5}{c}{\textbf{Moments}} \\ \hline
$b(p_\alpha)$ & $-2.8\times 10^{-5}$ & $-0.0011$  & $4.7\times 10^{-5}$ & $4.0 \times 10^{-6}$ \\
$e(p_\alpha)$ &0.065 &0.32  &0.089 & 0.011 \\ \hline
\multicolumn{5}{c}{\textbf{Peaks}} \\ \hline
$b(p_\alpha)$& 0.009 & 0.026 & $3.2\times 10^{-4}$ & 0.0016 \\
$e(p_\alpha)$ & 0.044  & 0.25  & 0.060 & 0.018 \\ \hline
\multicolumn{5}{c}{\textbf{Moments + Power spectrum}} \\ \hline
$b(p_\alpha)$& $3.2\times 10^{-5}$ & $-6.43\times 10^{-4}$ & $-3.38\times 10^{-5}$ & $7.24\times 10^{-6}$ \\
$e(p_\alpha)$ & 0.048  & 0.26  & 0.071 & 0.012 \\ \hline
\end{tabular}
\end{center}
\caption{Bias and marginalized errors (for a $3^\circ\times3^\circ$ field of view) on the parameters using different sets of descriptors. We used the 9 moments, 350 linearly spaced bins between $100\leq l \leq 2\times10^4$ for the power spectrum, 175 linearly spaced bins in $[-2\sigma,2\sigma ]$ for the MFs and 350 linearly spaced bins in $[-2\sigma,5\sigma]$ for the peaks. The maps were smoothed with a Gaussian smoothing kernel of scale $\theta_G=1^\prime$, and a single source plane at redshift $z=2$ was considered. A galaxy density $n(z=2)=15\,\mathrm{arcmin}^{-2}$ has been assumed.}
\label{errbiastable}
\end{table*}


\subsection{Robustness check: number of bins}
\label{robustness}
\begin{figure*}[htdp]
\begin{center}
\includegraphics[scale=0.4]{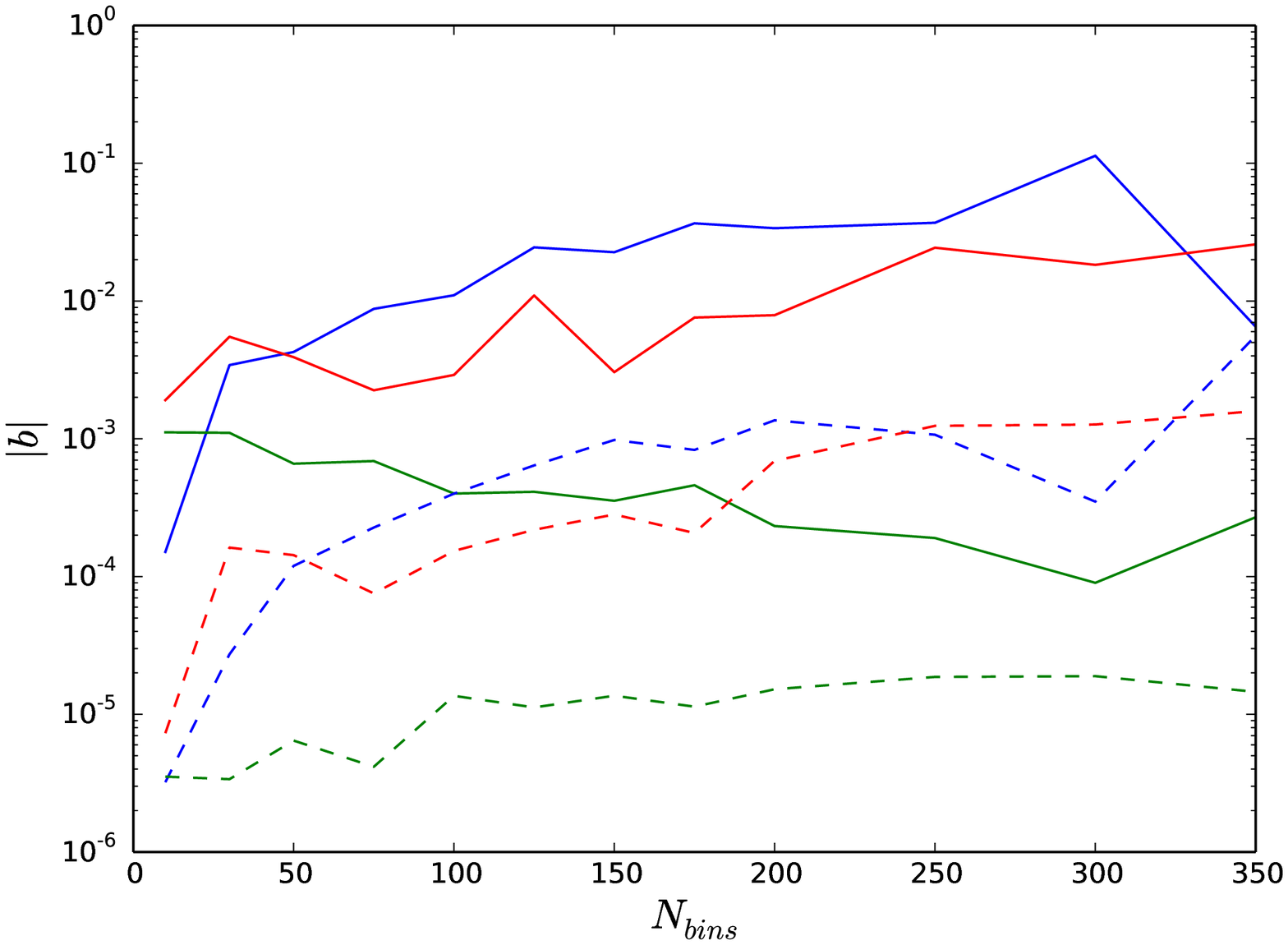}
\includegraphics[scale=0.4]{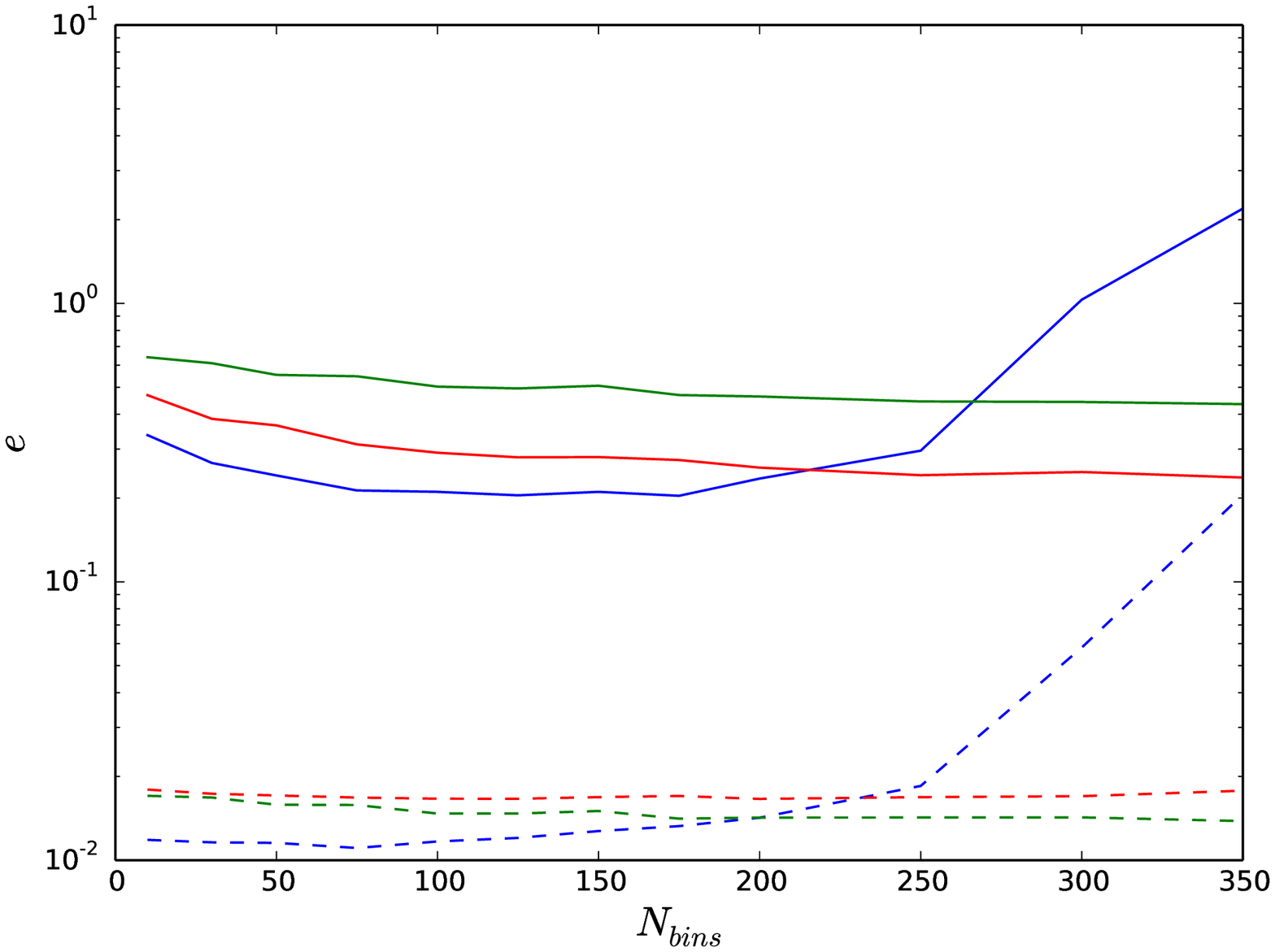}
\caption{Biases (left panel) and marginalized errors (right panel) on $w$ (solid) and $\sigma_8\Omega_m^{0.5}$ (dashed) as a function of $N_{bins}$ for the three statistics PS (green), MF (blue) and PK (red).}

\label{robustness1}
\end{center}
\end{figure*}

In this section we check how our results depend on the number of bins for the PS, MF and PK statistics. Note that when we specify $N_{bins} $ for MF, the observables vector is $3N_{bins}$ long because there are 3 MFs. The behavior of the biases $b(p_\alpha)$ and marginalized errors $e(p_\alpha)$ as a function of $N_{bins}$ for the parameters $(w,\sigma_8\Omega_m^{0.5})$ is shown in Figure \ref{robustness1}. The main conclusion we draw from this plot is that our results are stable and reach a plateau for $N_{bins}\approx 200$. For the the MF statistic they start to increase for a larger number of bins due to numerical instabilities. A similar behavior has already been observed in \citep{Petri2013}. 

\section{LSST spurious shear}
\label{lsstnoise}

\begin{figure*}
\begin{center}
\includegraphics[scale=0.8]{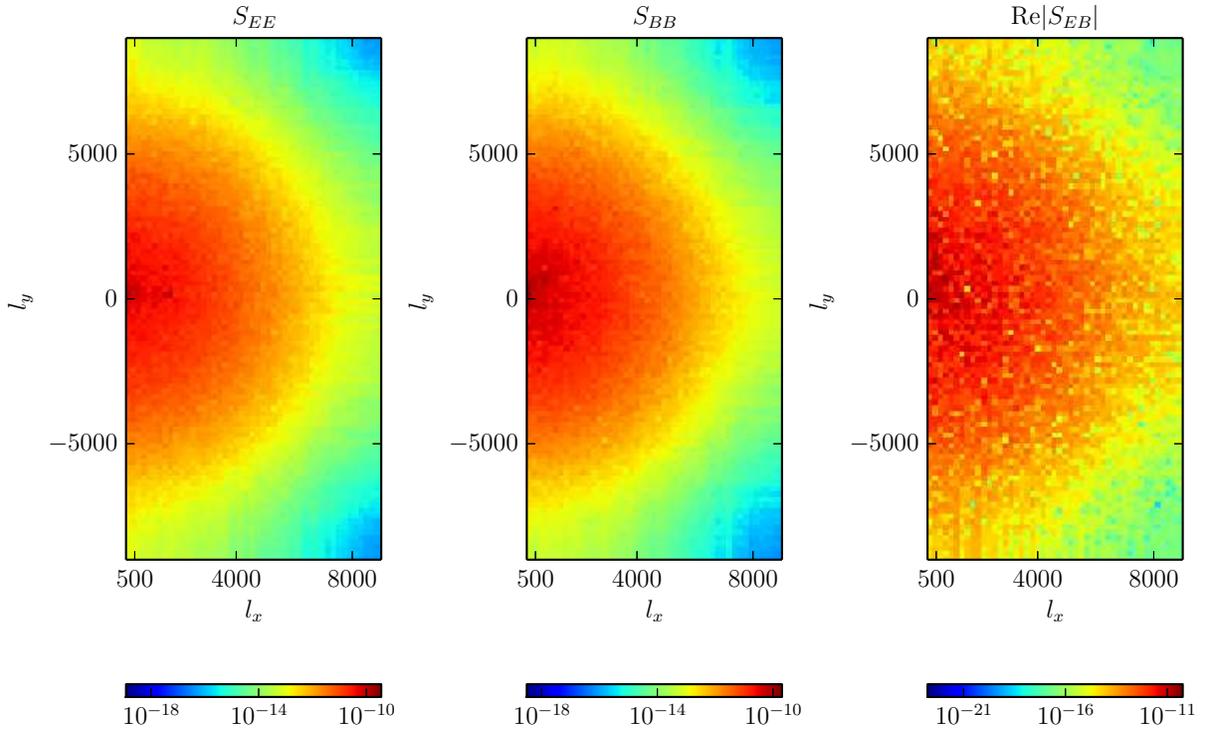}
\caption{$E$ and $B$ mode power spectral densities of the spurious shear measured from the LSST instrument simulations. The mode range $l\lesssim 10^4$ has been chosen to be lower than the scale parameter $1/\theta_G$ of the Fourier transformed Gaussian filter $\hat{W}_{\theta_G}(l)=\frac{\theta_G^2}{2\pi}\exp{\left(-\frac{l^2\theta_G^2}{2}\right)}$. Since the shear field is real, the negative $l_x$ semiplane does not contain additional information with respect to the positive $l_x$ plane.} 
\label{2dEB}
\end{center}
\end{figure*}

\citep{chihway} used an LSST instrument simulation (which relies on the extensive efforts of \citep{LSSTOperations}\footnote{See also \url{http://www.lsst.org/lsst/opsim} and \url{http://confluence.lsstcorp.org/display/PHOSIM}}) to calculate the spurious shear expected for LSST due to atmospheric effects, stochastic optics errors, tracking errors, and counting statistics. They used an LSST instrument specific code to simulate optical aberrations on galaxy shapes; their simulated fields of view are approximately $2^\circ \times 2^\circ$. The catalogs that we analyze describe the residual distortions after PSF corrections with polynomial fits. The stochastic piece of the spurious shear correlation decreases approximately with the inverse of the effective number of exposures with which each field of view is probed.  

We use the 20 spurious shear maps which  \citep{chihway}  generated to study the effect of spurious shear. These maps consist of a list of galaxies, with sky angular positions $\pmb{\theta}_i$ and additive spurious shear components $\gamma^{1,2}_i\equiv \gamma^{1,2}(\pmb{\theta}_i)$, where the shear components $\gamma^{1,2}$ are a measure of the residual ellipticity distortion that the atmosphere and instrument imprint on the galaxy images. \citep{chihway} analyzed these maps computing the two--point angular correlation function $\xi^+(\theta)$, defined as 
\begin{equation}
\label{xiplus}
\xi_{\gamma\gamma}^+(\theta)=\langle\gamma^1(\pmb{\theta}_i)\gamma^1(\pmb{\theta}_j)+\gamma^2(\pmb{\theta}_i)\gamma^2(\pmb{\theta}_j)\rangle,
\end{equation}
where the averaging is performed over all galaxy pairs separated by an angular distance $\vert\pmb{\theta}_i-\pmb{\theta}_j\vert=\theta$. Since the shear $\pmb{\gamma}$ is a two component spin 2 field, however, $\xi_{\gamma\gamma}^+$ alone does not characterize it completely, even in the Gaussian case. One can construct an independent shear two point function $\xi_{\gamma\gamma}^-$ with a different quadratic spin combination to recover the missing information content. 

Instead of using a real--space approach, we adopt the analogous $EB$ mode decomposition, which has been widely used to characterize the CMB polarization. This technique decomposes the Fourier--transformed shear field $\gamma^{1,2}(\mathbf{l})$ into its $E$ and $B$ components. Following \citep{BartelmannSchneider} we compute
\begin{equation}
\label{EBdecomposition}
\begin{matrix}
E(\mathbf{l}) = \left(\frac{l_x^2-l_y^2}{l_x^2+l_y^2}\right)\gamma^1(\mathbf{l}) + \left(\frac{2l_xl_y}{l_x^2+l_y^2}\right)\gamma^2(\mathbf{l}) \\
B(\mathbf{l}) = \left(\frac{-2l_xl_y}{l_x^2+l_y^2}\right)\gamma^1(\mathbf{l}) + \left(\frac{l_x^2-l_y^2}{l_x^2+l_y^2}\right)\gamma^2(\mathbf{l})
\end{matrix}
\end{equation} 
from which we calculate the power spectral densities $S_{EE},S_{BB}$ and the cross power $S_{EB}$. This decomposition is particularly useful since, for a pure lensing shear signal $P_{\kappa}=P_{EE}$, $P_{BB}=P_{EB}=0$ (note that we use the notation $P$ for the signal and $S$ for the spurious shear). Because of this, a non--null $B$ detection can be attributed to systematics leading to the possibility of correction. 

The $E$ and $B$ mode spectral densities contain the same information as the real space correlation functions $\xi_{\gamma\gamma}^{\pm}$, in particular
\begin{equation}
\label{correlationEB}
\xi_{\gamma\gamma}^+(\theta)=\int_0^{\infty}\frac{ldl}{2\pi}[S_{EE}(l)+S_{BB}(l)]J_0(l\theta)
\end{equation}
where $J_0$ is the zeroth order Bessel function of the first kind. Moreover in this fashion it is easy to model the additive convergence systematics, since $P_\kappa=P_{EE}$. 

We extract information about the $E$ and $B$ mode spurious spectral densities by analyzing the 20 simulated maps. For simplicity, we used the provided catalogs to construct pixelized shear maps $\gamma_P^{1,2}(\mathbf{p})$ (with $\mathbf{p}=(n_x\theta_{pix},n_y\theta_{pix})$, $n=1...N_{pix}$) using a pixelization smoothing procedure
\begin{equation}
\gamma_P^{1,2}(\mathbf{p})=\sum_\mathbf{q}\frac{\sum_i\gamma^{1,2}(\pmb{\theta}_i)\delta_{\mathbf{q},\pmb{\theta}_i}W_{\theta_G}(\vert\mathbf{p}-\mathbf{q}\vert)}{\sum_i\delta_{\mathbf{q},\pmb{\theta}_i}}
\end{equation}

%
\noindent where the first sum is over pixels, the second sum is over the galaxies in the catalog and the Kronecker $\delta$ symbol is defined as
\begin{equation}
\label{deltakron}
\delta_{\mathbf{p},\pmb{\theta}_i} = 
\begin{cases}
1, &\text{if } \pmb{\theta}_i \text{ falls in } \mathbf{p} \\
0, &\text{otherwise }
\end{cases}
\end{equation}
The smoothing kernel has been chosen to be a Gaussian,
\begin{equation}
W_{\theta_G}(\theta)=\frac{1}{2\pi\theta_G^2}\exp{\left(-\frac{\theta^2}{2\theta_G^2}\right)},
\end{equation}
with a scale parameter $\theta_G=1^\prime$; the pixel size has been chosen as $\theta_{pix}\approx0.2^\prime$, and each map has a total of $N_{pix}=512$ pixels per side (which corresponds roughly to a $2^\circ\times2^\circ$ field of view). The details of the smoothing procedure do not matter when we restrict ourselves to angular scales larger than $\theta_G=1^\prime$, or equivalently to $l$ modes smaller than $l_{max}\approx 2\times 10^4$. Using Fast Fourier Transforms (FFT), we measured an average over catalogs of the $E$ and $B$ modes power spectral densities. Figure \ref{2dEB} shows that these are consistent with a statistically isotropic spurious shear, with equal amount of power in the $E$ and $B$ channels and a weak $EB$ correlation. We quantified this correlation measuring the correlation coefficient $\gamma_{EB}(5000)\approx 0.1$ where
\begin{equation}
\gamma_{EB}(l) = \frac{P_{EB}(l)}{\sqrt{P_{EE}(l)P_{BB}(l)}}.
\end{equation} 
%
\begin{figure*}
\begin{center}
\includegraphics[scale=0.4]{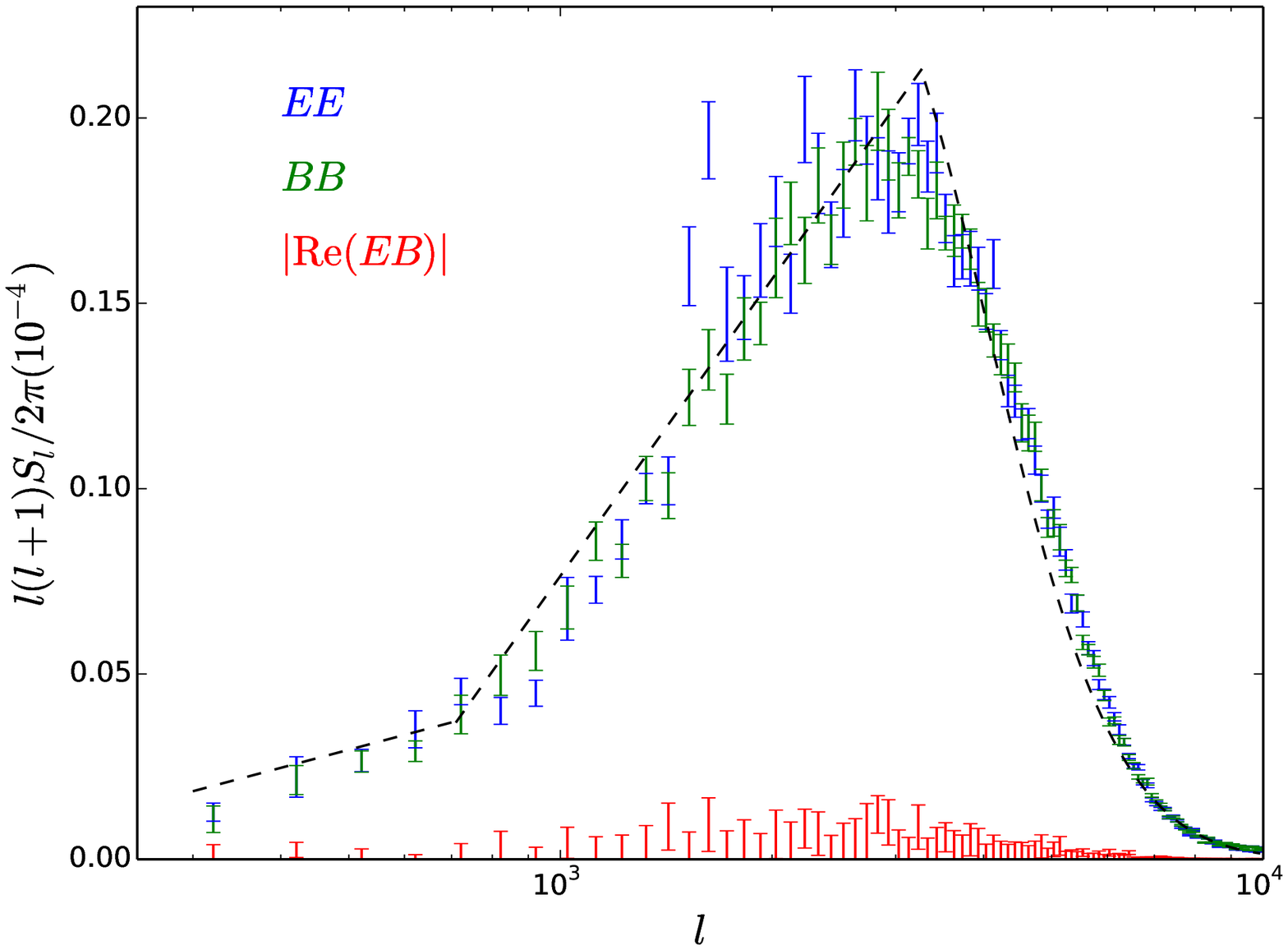}
\includegraphics[scale=0.43]{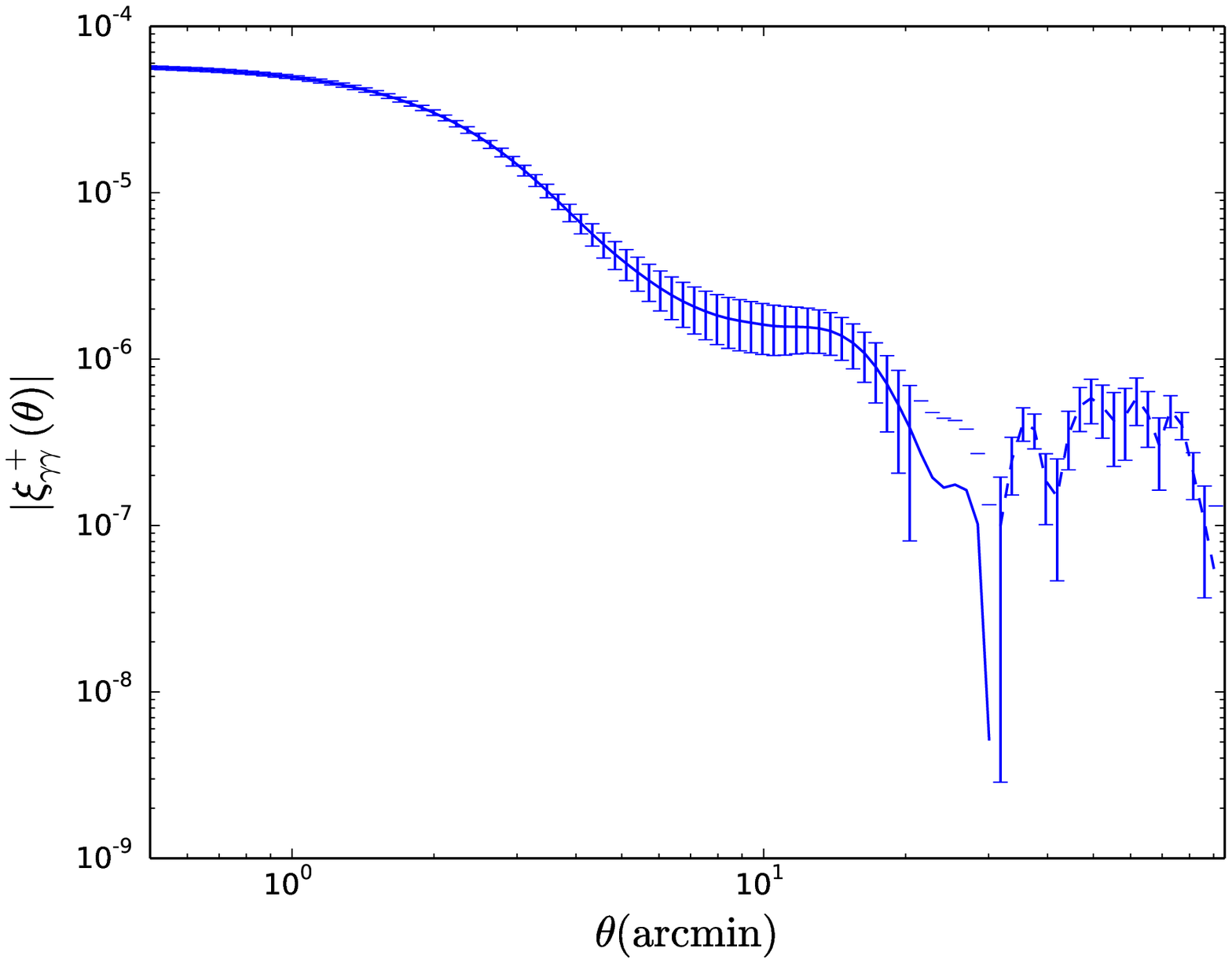}
\caption{Left panel: the $E$ and $B$ mode power spectral densities of the spurious shear measured from the LSST instrument simulation, $S_{EE}$ (blue), $S_{BB}$(green), $\vert\mathrm{Re} S_{EB}\vert$ (red). We fitted the parameters $(A_0,n_0,A_1,n_1,A_2,b,\mu)$ as in equation (\ref{improvedloglin}) (keeping $l_0=700.0$ fixed) to the measured $E$ power spectrum profile; the black dashed line corresponds to the best fit model parameters. Right panel: spurious shear angular correlation function $\xi^+_{\gamma\gamma}(\theta)$ calculated from the $E$ and $B$ power spectra as in (\ref{correlationEB}); the solid and dashed lines refer to positive and negative values respectively.} 
\label{EBpow1d}
\end{center}
\end{figure*}
We also plot the individual components of the $E$ and $B$ mode power spectra in Figure \ref{EBpow1d}. We find that the log-linear model in equation (\ref{logpower}) is not a good description of the measured $E$ mode power spectrum, failing both at small and large $l$. In particular, we believe that the excess feature that appears at $\theta\lesssim 3^\prime$ in the angular correlation function in Figure \ref{EBpow1d} is due to the damping of the spurious shear power spectrum at high $l$. The spurious shear excess on $\lesssim3$\,arcmin scales can be attributed to the inability of the polynomial fits to capture the variation of the PSF on small scales. Because of this our results, which include the treatment of this larger small-scale noise, are very conservative. \citep{chihway} do not reproduce this excess feature with the log-linear model since this model is scale free and decreases too slowly at high $l$. 

We introduced a damping scale in the model through an analytical description of the measured $E$ mode spurious shear power spectrum that is piecewise log-linear for $l\leq3300$ and has an exponential damping for $l>3300$.
\begin{equation}
\label{improvedloglin}
S_l = 
\begin{cases}
\frac{A_0}{l(l+1)}\left[n_0\log{\left(\frac{l}{l_0}\right)+1}\right], &\text{if } 0<l\leq700 \\
\frac{A_1}{l(l+1)}\left[n_1\log{\left(\frac{l}{l_0}\right)+1}\right], &\text{if } 700<l\leq3300 \\
\frac{A_2\log{l}}{l(l+1)}\exp{\left[-b(\log{l}-\mu)^2\right]}, &\text{if } l\geq3300 
\end{cases}
\end{equation}

We found the best fit parameters for this model to be $(A_0,n_0,A_1,n_1,A_2,b,\mu)=(3.17\times 10^{-5},1.36,1.6\times10^{-4},7.54,4.4\times 10^{-5},15.37,3.41)$. We also quantified the order of magnitude of the spurious shear squared amplitude as $\sigma_{sys}^2\approx 3.6\times 10^{-5}$. This reduces to $\sigma^2_{sys}\approx 10^{-7}$ once we divide it by the number of exposures $N_{exposures}=368$. 

The fitting formula we obtained provides an accurate fit to the power spectrum of the spurious shear, and can be used to investigate the impact of the spurious shear. However, we caution that its parametrization is ad-hoc, and it cannot be used directly in obtaining Bayesian best fits to data (see \citep{astroMLText}). 

Since we want to study the scale--dependence of the cosmological parameter biases, we use a linear interpolation of the measured $E$ power spectrum in Figure \ref{EBpow1d} to generate our Gaussian mock spurious shear maps. The damping of the noise at high $l$ can in principle have an effect on the bias values we obtain in \S~\ref{results}. 
\begin{table*}
\begin{center}
\begin{tabular}{cccc} 

\multicolumn{4}{c}{\textbf{Survey Assumptions 4}} \\
\multicolumn{4}{c}{$n(z)=n_g\delta(z-2)$, $n_g=15\,\mathrm{arcmin}^{-2}$, $l\in[100,2\times10^4]$, $\nu^{MF}\in[-2,2]$, $\nu^{pk}\in[-2,5]$} \\ \hline

& $\Omega_m$ & $w$ & $\sigma_8$  \\ \hline
\multicolumn{4}{c}{\textbf{Power spectrum}} \\ \hline
Log-linear & $4.0\times 10^{-6}$  & $-2.69\times 10^{-4}$ & $2.5\times 10^{-5}$ \\
LSST simulation &  $-6.22\times10^{-5}$ &  $2.94\times10^{-4}$ &  $1.32\times10^{-4}$ \\
LSST simulation $\times 10$ & $-7.51\times10^{-4}$ &  0.0025 &  0.0015 \\ \hline

\multicolumn{4}{c}{\textbf{Minkowski Functionals}} \\ \hline
Log-linear & 0.0026 &0.037 & $-0.0024$ \\
LSST simulation & 0.0020 &  0.025 & $-0.0014$ \\
LSST simulation $\times 10$ & 0.007 & 0.055 & $-0.0068$ \\ \hline

\multicolumn{4}{c}{\textbf{Moments}} \\ \hline
Log-linear & $-2.8\times 10^{-5}$ & $-0.0011$  & $4.7\times 10^{-5}$ \\
LSST simulation & $1.09\times10^{-5}$ & $-3.96\times10^{-4}$ & $-7.60\times10^{-6}$ \\
LSST simulatiion $\times 10$ & $-2.84\times10^{-5}$ & $-4.72\times10^{-3}$ &  $1.26\times10^{-4}$ \\ \hline

\multicolumn{4}{c}{\textbf{Peaks}} \\ \hline
Log-linear & 0.009 & 0.026 & $3.2\times 10^{-4}$ \\
LSST simulation & 0.0011 &  0.018 &  $2.9\times10^{-4}$ \\
LSST simulation $\times 10$ & 0.0026 & 0.046 & $4.0\times10^{-4}$ \\ \hline
\end{tabular}
\end{center}
\caption{Comparison for the bias values on the parameter triplet $(\Omega_m,w,\sigma_8)$ using three different models for the LSST spurious shear: ``Log-linear" refers to the log-linear model with $(A,n,l_0)=(10^{-6.6},0.7,700)$, with the normalization $\sigma^2_{sys}=4\times10^{-7}$, repeated from Table IV. ``LSST simulation" refers to the power spectrum measured from the maps of \citep{chihway} divided by $N_{exposures}=368$,
 ``LSST simulation $\times$ 10" refers to the same model but with the amplitude $\sigma^2_{sys}$ increased by a factor of 10.}
\label{comparebiasesmodels}
\end{table*}
In Table \ref{comparebiasesmodels} we see that the biases induced by spurious shear calculated from the LSST instrument simulation maps (``LSST simulation") for the morphological statistics (MF,PK) are smaller 
than for the ``Log-linear" model of Table \ref{errbiastable}. This is a result of two opposing factors. The amplitude of $\sigma_{sys}^2$ measured from the maps, scaled with the number of exposures, is smaller than assumed in \ref{errbiastable}. 
However, the damped noise power spectral shape (\ref{improvedloglin}) has more power on large scales, $l\lesssim 3300$, with respect to the simple log-linear one, and the bias on the parameters seems to come mainly from this large scale component. We also see that if we increase  $\sigma_{sys}^2$ by a factor of $10$, the biases obtained with the PS statistics scale linearly as expected, while deviations from a simple linear scaling are observed for the other statistics.
The bias in the morphological statistics, though small, needs to be further reduced for surveys with the statistical power of LSST. In the future, in addition to already described analysis enhancements, we could use the fact that the spurious shear power in the $E$ and $B$ are similar, to correct for these sources of contamination in the signal. Treating the noise as a Gaussian random field with known power spectral shape, maximum likelihood denoising procedures on the convergence maps become possible (see \citep{Descart} for an example of maximum likelihood denoising of CMB polarization maps). 
In Table \ref{momentsconnected}, we investigate the non-Gaussianities
in the spurious shear maps, measuring a set of 9 cubic and quartic
moments of the spurious shear $E$ mode. The first column in the Table
shows the values of the skewness and kurtosis moments, averaged over
the twenty independent $2\times2$deg$^2$ LSST noise maps.  The second
column shows the equivalent quantities measured from mock Gaussian
maps.  The results show that non-Gaussianities in the spurious shear
are not small, and in particular the kurtosis moments measured from
the spurious shear maps are much larger than those introduced in our
mock (Gaussian) spurious shear realizations. Nevertheless, for a survey with a size
comparable to the cumulative area of the twenty LSST noise maps
($80$deg$^2$), we do not expect these non-Gaussianities in the
spurious shear to affect our results significantly.  This is because
the skewness and the kurtosis are still much smaller than the
non-Gaussianities in the lensing signal. Furthermore the non-Gaussianity in
the skewness is modest, while the the kurtosis moments do not add
significant cosmological information to that already captured by the
skewness moments (see again \citep{Petri2013}).  However, the scaling
of this conclusion to larger surveys is not possible without computing the
non-Gaussianities of the LSST noise in correspondingly larger maps.
\begin{table*}
\begin{center}
\begin{tabular}{c|c|c|c} 
Moment & Value (20 LSST realizations) & Value (20 gaussian mock realizations) & Value (20 lensing signal realizations) \\ \hline
$S_0$ & $1.68\times10^{-2}$ & 5.8$\times 10^{-3}$  & 1.0 \\
$S_1$ & $3.67\times10^{-3}$ & $-$3.2$\times 10^{-3}$  & $-$1.4  \\
$S_2$ & $1.87\times10^{-2}$ & 2.5$\times 10^{-3}$  & $-$0.57  \\ \hline
$K^c_0$ & $1.38\times10^{-1}$ & $-$1.3$\times 10^{-2}$  & 2.7  \\
$K^c_1$ & $-6.91\times10^{-1}$ & 8.3$\times 10^{-3}$ & $-$4.4 \\
$K^c_2$ & $-4.55$ & 6.9$\times 10^{-4}$  & $-$78.0  \\
$K^c_3$ & $7.48$ & $-$3.2$\times 10^{-3}$ & 150.0 \\
\end{tabular}
\end{center}
\caption{Values of cubic $(S)$ and quartic $(K)$ moments (see \citep{Petri2013} for the precise definitions) measured from spurious shear maps, from LSST simulations \citep{chihway}, our mock Gaussian spurious shear maps, and from the actual lensing signal from IGS1 simulations.}
\label{momentsconnected}
\end{table*}
%


\section{Discussion}
\label{discussion}

\begin{table*}
\begin{center}
\begin{tabular}{ccccc}

\multicolumn{5}{c}{\textbf{Survey Assumptions 5}} \\
\multicolumn{5}{c}{$n(z)=n_g(\delta(z-1) + \delta(z-2))$, $n_g=15\,\mathrm{arcmin}^{-2}$, $l\in[100,2\times10^4]$, $\nu^{MF}\in[-2,2]$, $\nu^{pk}\in[-2,5]$} \\ \hline

& $\Omega_m$ & $w$ & $\sigma_8$ & $\sigma_8\Omega_m^{0.5}$ \\ \hline
\multicolumn{5}{c}{\textbf{Power spectrum}} \\ \hline
$b(p_\alpha)$ & $5.2\times 10^{-5}$  & $-1.65\times 10^{-4}$ & $1.0\times 10^{-4}$ & $9.2\times 10^{-5}$ \\ 
$e(p_\alpha)$ & 0.028 & 0.31 & 0.042 & 0.010 \\ \hline
\multicolumn{5}{c}{\textbf{Minkowski}} \\ \hline
 $b(p_\alpha)$ &0.0025 &0.022 &-0.0014 & 0.0011  \\
 $e(p_\alpha)$ &0.035 &0.21  &0.047 & 0.012\\ \hline
\multicolumn{5}{c}{\textbf{Moments}} \\ \hline
$b(p_\alpha)$ & $1.1\times 10^{-4}$ & 1.6$\times 10^{-4}$  & $-1.8\times 10^{-4}$ & $-2.95\times10^{-5}$ \\
$e(p_\alpha)$ &0.037 &0.26  &0.044 & 0.0093 \\ \hline
\multicolumn{5}{c}{\textbf{Peaks}} \\ \hline
$b(p_\alpha)$& 0.0017 & 0.022 & -0.0011 & $7.3\times 10^{-4}$ \\
$e(p_\alpha)$ & 0.039  & 0.23  & 0.054 & 0.017 \\ \hline
\multicolumn{5}{c}{\textbf{Moments + Power spectrum}} \\ \hline
$b(p_\alpha)$& $3.81\times 10^{-5}$ & $-9.22\times 10^{-5}$ & $-9.36\times 10^{-5}$ & $-2.03\times10^{-5}$  \\
$e(p_\alpha)$ & 0.027  & 0.22  & 0.038 & 0.0093 \\ \hline
\end{tabular}
\end{center}
\caption{Same as Table \ref{errbiastable}, but with redshift tomography using two redshifts $(z_{s,1},z_{s,2})=(1,2)$; a galaxy angluar density $n(z_{s,1})=n(z_{s,2})=15\,\text{arcmin}^{-2}$ has been assumed}
\label{errbiastabletomo}
\end{table*}
\begin{table}
\begin{center}
\begin{tabular}{cccc}
& $\Omega_m$ & $w$ & $\sigma_8$  \\ \hline
\multicolumn{4}{c}{\textbf{Minkowski}} \\ \hline
Bias & 0.0011 &  0.015 & $-6.0\times10^{-4}$ \\
Error & 0.052 & 0.26 & 0.076 \\ \hline
\multicolumn{4}{c}{\textbf{Peaks}} \\ \hline
Bias & $4.0\times10^{-4}$ & 0.021 & 0.0012 \\
Error & 0.055 &  0.28 &  0.069 \\ \hline
\end{tabular}
\end{center}
\caption{Biases and marginalized errors on the $(\Omega_m,w,\sigma_8)$ triplet calculated considering only peaks with $\nu>\nu^{pk}_{m}=1$ and excursion sets with $\nu>\nu_{m}^{MF}=1$ for Minkowski functionals}
\label{thresholdraise}
\end{table}

The results presented above can be divided into three main parts, which we now discuss in turn.
In the first part, we estimate the constraining power and associated bias on the cosmological parameters using only the convergence power spectrum statistic, which we calculate using the public code NICAEA \citep{nicaea}. In this case we do not have a strict limit on the cosmological parameters we can consider and hence have the freedom to consider the full set of $\Lambda$CDM cosmological parameters (in either of the two parametrizations $p^{1,2}_\alpha$), and to consider arbitrary variations of their numerical values. This freedom allows us to perform stability checks on our Fisher analysis, such as varying the stepsizes for the finite difference derivatives. When we adopt an iterative approach to determine the optimal stepsize, the results converge after few iterations; these results are a factor of 1.5 different compared to taking finite differences using fixed 20\% stepsizes. 


We are also able to compare some of our results on bias and marginalized constraints with previous published work. We reproduced exactly the initial conditions and assumptions in \citep{amara} and, by performing the same calculations, we find somewhat different values both for the biases and marginalized constraints (which we list in Table \ref{bias7parTable} bottom). In particular we find that our calculated biases are a factor of $\sim$10 smaller that the ones in \citep{amara}.

The second part of this work focuses on bridging the gap between a NICAEA-like semi-analytical approach, and a full numerical one, using simulated convergence maps which were created from a suite of ray--tracing N-body simulations. The IGS1 set of simulated maps is limited in the cosmological parameters we are able to vary, i.e. only the $(\Omega_m,w,\sigma_8)$ triplet, by the step sizes for the finite difference derivatives which are of order of 20\%, and by the underlying galaxy distribution assumed in ray--tracing. We limit ourselves to the case where all the source galaxies lie on a single redshift plane at $z_s=2$. The simulated fields of view are approximately $3^\circ\times3^\circ$, with a pixel resolution of $0.1^\prime$, which limits us to to an $l$ range of $[10^2,2\times10^4]$, with $l$ bins which must be sized at least $\delta l_{bin}\sim 100$. 

In order to compare the results of the simulations to the ones obtained with NICAEA, in practice we restrict ourselves to the $l$ range $[500,5000]$ to avoid both the large--scale modes that have little constraining power, and the heavily nonlinear small--scale modes. Table \ref{comptable} shows a comparison between the marginalized errors and biases obtained, under the same conditions, with a NICAEA semi-analytical approach and with fully numerical methods. We immediately see that there are some discrepancies in the $(\Omega_m,\sigma_8)$ marginalized constraints, which we attribute to limitations in the simulations. In particular Figure \ref{derivatives} shows the presence of numerical inaccuracies in the power spectrum derivatives measured from the simulated maps. These inaccuracies have a large impact on the $(\Omega_m,\sigma_8)$ doublet, because of its intrinsic degeneracy. We tried to mitigate the effect of this degeneracy by restricting our forecasts to the $\sigma_8\Omega_m^{0.5}$ parameter combination, which corresponds to the minimum variance direction in the $(\Omega_m,\sigma_8)$ likelihood plane; the effect of this mitigation can be seen in the last column of Table \ref{comptable}, which shows better agreement between the NICAEA and fully numerical approaches. 

In the third part of this work, we estimate the impact of spurious shear on the nonlinear statistics that we consider in addition to the power spectrum: moments (LM), peak counts (PK) and Minkowski functionals (MF). Because of the complicated analytical structure of these statistics, a NICAEA--like semi-analytical approach is not yet possible, though it might be in the future as emulation software is being developed. For now we are forced to restrict ourselves to the IGS1 simulated maps. Table \ref{errbiastable} shows the effect of the spurious shear on the three non--linear statistics.
Constraints derived from the moments and from the power spectrum have bias $\ll$ marginalized errors for $(w, \sigma_8 \Omega_m^{0.5})$.   In the ideal case that the spurious noise remained Gaussian, and the biases were
independent of survey size, they would remain negligible even for a survey with the statistical power of LSST
(scaling the errors by a factor of $1/40$).  However, as discussed above, this scaling is unlikely to hold and needs to be verified by quantifying non-Gaussianities in the noise in larger noise maps.

When the moments are combined with the power spectrum, the constraints are tightened by almost a factor of 2. However, the morphogical statistics (MF,PK) suffer from important biases, which need to be corrected for before applying to surveys of the statistical power of LSST. 

Redshift tomography may reduce bias, since the spurious shear is a constant, redshift independent, addition to the shear which cannot easily mimic the redshift dependence of the true cosmological lensing signal.  We find, for example, that combining only two redshifts (which are the only ones available in the IGS1 simulation suite) decreases the bias of
$\sigma_8 \Omega_m^{0.5}$
calculated from peaks by a factor of $\sim$2, accompanied by a ~10\% decrease in the bias of $w$. This redshift combination, however, shrinks the errors too making the ratio $b/e$ worse in some cases, as shown in Table \ref{errbiastabletomo}. Another possible way to reduce bias is to restrict the application of the morphological statistics to higher peaks, $\kappa>\nu^{pk}_m\sigma_0$, and higher excursion sets $\Sigma=\{\kappa>\nu^{MF}_m\sigma_0\}$ for the Minkowski Functionals, where one can hope that the spurious shear does not have important effects. Such investigation has been performed in Table \ref{thresholdraise}, which shows that some improvement in the $\vert b/e\vert$ ratio can be obtained by imposing a modest low--threshold $\nu_m$.   

Including a more complete set of tomographic bins, additional
cosmological parameters, Planck priors and simultaneous fitting MF,PK,
LM and PK constraints along with self--calibration must be addressed in future work, but is not
possible with the limited simulation set we are working with. New simulations with a more realistic galaxy redshift distribution and a much larger number of realizations to permit simultaneous analysis of multiple statistics and redshifts are required. 

We performed a robustness check to make sure our results are free from numerical instabilities which mainly arise from not having enough realizations to estimate a too large covariance matrix. This robustness check is distinct from the ones we performed in the comparison between NICAEA and the simulations. In those comparisons we kept the number of bins fixed and studied the simulation inaccuracies; this robustness check on the other hand shows that with too many bins the covariance matrix becomes singular and the resulting errors become bigger than the ones that those due to a misestimation of the derivatives. 

Figure \ref{robustness1} suggests an optimal number of bins to adopt, so that the marginalized errors reach a plateau and do not blow up due to the covariance matrix becoming singular for a too large $N_{bins}$. A similar behavior has already been observed in \citep{Petri2013}. 

As an interesting and timely application of our methods, we explored their implications for the published \citep{chihway} simulations of LSST data. 
We determined the biases in cosmological parameters implied by the published spurious shear. Analyzing the statistical properties of the LSST simulated spurious shear, we found that its skewness is very small ($O(10^{-3})$), and hence we do not expect it to affect our conclusions on the parameter biases.One should note that this value refers to a simulated sky coverage of $\sim 80\mathrm{deg}^2$, and for a larger survey the noise non-Gaussianity is not known yet. Since this could affect our conclusions, this will need investigated when noise  simulations covering larger areas become available.

We performed an $E/B$ mode decomposition of the LSST simulated spurious shear, and found that the amount of power in the $E$ and $B$ channels is very similar, and we hope that this result can be used in the future to correct for this kind of systematic. One should note, however, that the equations in (\ref{EBdecomposition}) are only exact for full sky coverage, and corrections have to be applied on the small $l$ multipoles due to the finite field of view of the simulations. Nevertheless, we do not expect our main conclusion, namely $P_{EE}\sim P_{BB}$ to be affected by these window effects; moreover, we take into account the increased variance of the small multipoles in our cosmological parameter inferences. We also find an excess in the small angle spurious shear correlation function, with respect to the assumed log--linear model, which can be mitigated in the future by more accurate PSF modeling. Accuracy of PSF modeling has already proved to be an important issue when analyzing existing weak lensing surveys, such as CFHTLens \citep{Kilbinger2013}, in which $\sim 25\%$ of the dataset had to be removed due to PSF contamination of the galaxy shear; removing this data resulted in
changes to the shear correlation function of nearly a factor of $\sim 2$ on large scales.


\section{Conclusions}
\label{conclusions}

%
%
%
%
%
%

In this work we investigate the effects of spurious correlated shear errors on weak lensing statistics, analyzing their effects on the four cosmological probes (PS,MF,LM,PK) in ray--tracing simulations in a unified fashion. Important pioneering work in this direction has already been done by \citep{DebbiePeaks} 
and by \cite{Shirasaki+2013,ShirasakiYoshida2014}.
Our main goal here is to have a full comparative analysis of the effect of spurious shear on all our cosmological probes, both the Gaussian (power spectrum) and non-Gaussian ones (moments, peak counts and Minkowski functionals). This paper should be considered a first step in this sense, and the main results of this analysis can be summarized as follows: 
\begin{itemize}
\item Using the power spectrum code NICAEA, we were able to calculate the Fisher forecasts on biases and errors on a set of 7 $\Lambda$CDM cosmological parameters; we found that, assuming the same level of spurious shear, the biases on the parameters are a factor of 10 smaller than found in previous work \citep{amara}. 
\item Assuming a log--linear spurious shear described by the power spectral shape in equation (\ref{logpower}), the power spectrum (PS) and moments (LM) give less biased parameter fits than the morphological descriptors (MF,PK). However, non-Gaussianity in the noise (not included in our analysis) will likely be important and must be quantified to assess the biases.
\item Our results highlight the need for more extensive lensing simulations and more accurate spurious shear estimates. A possible theoretical improvement that could greatly help our investigation is the development of emulation software (NICAEA-like) that would allow a semi--analytical treatment of the nonlinear statistics. We hope to address this need in the future. 
\end{itemize} 

\section*{Acknowledgements}
We thank Chihway Chang for providing us with the LSST spurious shear
maps, and for many useful related discussions. We thank the LSST simulation effort, whose work made possible the calculations of the spurious shear expected in LSST data in \citep{chihway}. We are also grateful to Hu Zhan for discussions of his calculations of the LSST figure of
merit and treatment of systematic errors.  This research utilized resources at the New York Center for
Computational Sciences, a cooperative effort between Brookhaven
National Laboratory and Stony Brook University, supported in part by
the State of New York. This work is supported in part by the
U.S. Department of Energy under Contract No. DE-AC02-98CH10886 and by
the NSF under grant AST-1210877 to ZH.  The simulations were created on the
IBM Blue Gene/L and /P New York Blue computer and the maps were
created analyzed on the LSST/Astro Linux cluster at BNL.
\bibliography{ref}
\label{lastpage}
\end{document}